\documentclass[10pt,journal,compsoc,onecolumn]{IEEEtran}
\usepackage[nocompress]{cite}
\usepackage{caption}
\usepackage{graphicx}
\usepackage{textcomp}
\usepackage{array}
\usepackage{lineno}
\usepackage[hidelinks]{hyperref}
\usepackage[linesnumbered,ruled]{algorithm2e}
\usepackage{amssymb}
\usepackage{amsmath}
\usepackage{multirow}
\usepackage{lscape}
\usepackage{epstopdf}
\modulolinenumbers[5]

\def \beqi{\begin{IEEEeqnarray}{rcl}\IEEEyesnumber}
\def \eeqi{\end{IEEEeqnarray}}

\def \beq  { \begin{equation} }
\def \eeq { \end{equation} }
\def \beqn{ \begin{eqnarray} }
\def \eeqn{ \end{eqnarray} }
\def \bmat{\begin{bmatrix}}
\def \emat{\end{bmatrix}}
\def \bmats{\left[\begin{smallmatrix}}
\def \emats{\end{smallmatrix}\right]}

\def \be {\begin{eqnarray}}
\def \ee {\end{eqnarray}}
\def \ben {\begin{eqnarray*}}
\def \een {\end{eqnarray*}}

\begin{document}
\sloppy

\title{Clustering-Based Collaborative Filtering Using an Incentivized/Penalized User Model}
\author{{Cong Tran},~\IEEEmembership{Student Member,~IEEE}, {Jang-Young Kim},~{Won-Yong Shin},~\IEEEmembership{Senior Member,~IEEE}, {and Sang-Wook Kim}
\IEEEcompsocitemizethanks{\IEEEcompsocthanksitem C. Tran is with Department of Computer Science and Engineering, Dankook University, Yongin 16890, Republic of Korea and the Department of Computational Science and Engineering, Yonsei University, Seoul 03722, Republic of Korea.\protect\\
Email: congtran@ieee.org. 
\IEEEcompsocthanksitem J.-Y. Kim is with the Department of Computer Science, University of Suwon, Hwaseong 18323, Republic of Korea.\protect\\
E-mail: jykim77@suwon.ac.kr.
\IEEEcompsocthanksitem W.-Y. Shin is with the Department of Computational Science and Engineering, Yonsei University, Seoul 03722, Republic of Korea.\protect\\
E-mail: wy.shin@yonsei.ac.kr.
\IEEEcompsocthanksitem S.-W. Kim is with the Department of Computer Science and Engineering, Hanyang University, Seoul 04763, Republic of Korea.\protect\\
E-mail: wook@agape.hanyang.ac.kr.
 }
}

\IEEEtitleabstractindextext{
\begin{abstract}
Giving or recommending appropriate content based on the quality of experience is the
most important and challenging issue in recommender systems. 
As collaborative filtering (CF) is one of the most prominent and popular techniques used for recommender systems, we propose a new \textit{clustering-based} CF (CBCF) method using an incentivized/penalized user (IPU) model only with ratings given by users, which is thus easy to implement.
We aim to design such a simple clustering-based approach with no further prior information while improving the recommendation accuracy. To be precise, the purpose of CBCF with the IPU model is to improve recommendation performance such as \textit{precision}, \textit{recall},
and $F_1$ score by carefully exploiting different preferences among users.
Specifically, we formulate a constrained optimization problem, in which we aim to
maximize the \textit{recall} (or equivalently $F_1$ score) for a given \textit{precision}.
To this end, users are divided into several clusters based on the
actual rating data and Pearson correlation coefficient. Afterwards, we give each
item an \textit{incentive/penalty} according to the preference tendency by users within the
same cluster.
Our experimental results show a significant
performance improvement over the baseline CF scheme without clustering in terms of \textit{recall} or $F_1$ score for a given \textit{precision}.
\end{abstract}

}

\maketitle
\IEEEdisplaynotcompsoctitleabstractindextext

%
\IEEEpeerreviewmaketitle
\section{Introduction}
\label{Introduction}

People are likely to have an increasing difficulty in finding their favorite content effectively since extensive collections of video, audio, papers, art, etc. have been created both online and offline.
For example, over hundreds of feature films and hundreds of thousands of books
have been produced and published every year in the US. However, one person would read at most about 10,000 books in his/her life, and then he/she must choose his/her favorite books among them.
On the one hand, recommender systems have been developed and used in diverse domains (e.g., the movie industry, the music
industry, and so on) by helping people to select appropriate content based on individual preferences~\cite{Adomavicius05}.
Especially, online commerce
industries such as Amazon.com and Netflix have successfully exploited how to increase
customer loyalty.
For example, Amazon.com and Netflix have generated much of their sales by providing personalized items
through their own recommender systems~\cite{Linden03, Koren09}.

While diverse recommender systems such as personalized recommendations, content-based recommendations,
and knowledge-based recommendations have been developed, collaborative filtering (CF) is one of the most prominent and popular techniques used for recommender
systems~\cite{Konstan12,Su09}.
CF methods are generally classified into memory-based CF and model-based CF.
In model-based CF, training datasets are used to develop a model
for predicting user preferences. 
Different machine learning techniques such as Bayesian networks, clustering, and rule-based approaches
can also be utilized to build models. An alternating least squares with weighted $\lambda$-regularization (ALS-WR) scheme is a representative example of model-based CF. ALS-WR is performed based on a matrix factorization algorithm and is tolerant of the data sparsity and scalability~\cite{Zhou08, Hu08}.
The main advantages of model-based CF are an improvement of prediction performance and
the robustness against the data sparsity.
However, it has some shortcomings such as an expensive cost for building a
model~\cite{Su09}.
On the other hand, memory-based CF does not build a specific model,
but directly computes the similarity between users or items using the entire rating matrix or its samples.
Hence, memory-based CF is easy to implement and effective to manage.
However, it has also some drawbacks such as dependence on human ratings,
performance decrement when data are sparse, and disability of recommendation for new users (i.e., cold-start users)
and items~\cite{Su09}.

Memory-based CF approaches are again classified into user-based CF and item-based CF.
The main ideas behind the user-based CF and item-based CF approaches are to find the
user similarity and the item similarity, respectively, according to the ratings (or preferences).
After finding similar users, called $neighbors$, user-based CF
recommends the top-$N$ most preferable items that an active user has not accessed yet.
User-based CF has limitations related to scalability, especially when
the number of users is much larger than the number of items.
Item-based CF was proposed to mitigate this scalability problem, but cannot
still entirely solve the problem when the numbers of users and items
are large.
Despite such limitations, CF has been employed as one of the most representative recommender systems leveraged in online commerce.

In addition, there have been many studies on the design of CF algorithms in terms of reducing the mean absolute error (MAE) or root mean squared error (RMSE) of rating prediction~\cite{Cai14}. However, recommender systems designed in the sense of minimizing the MAE or RMSE do not inherently improve recommendation accuracy.
We assume that there are two recommender systems having the same MAE or
RMSE of the rating prediction.
We note that they may differ from each other in terms of user experience (UX) since
there is a possibility that one recommender system recommends an item whereas the other does not.
For example, suppose that the real preference of a user on an item is 4.2 and two recommender systems
predict the preference as 3.8 and 4.6, respectively.
Then, when items having the predicted preference of more than 4.0 are assumed to be recommended, the MAEs of two recommender systems are the same but only the latter one will recommend the
item.
In order to redeem the above case, some performance metrics related to UX
such as $precision$, $recall$, and $F_1$ score have been widely used in the literature.

On the other hand, several companies, e.g., Pandora Internet Radio, Netflix, and Artsy,
have developed their own clustering-based recommendation methods, called Music Genome Project,
Micro-Genres of Movies, and Art Genome Project, respectively.
These clustering-based recommendation methods have successfully led to satisfactory performance,
but the processing cost for clustering is very expensive.
For example, it is widely known that each song tends to be analyzed by a musician through a process that
takes usually 20 to 30 minutes per song in the case of Music Genome Project.

Unlike the aforementioned clustering-based recommendation methods that take long processing time to recommend items, we aim to design a simple but novel \textit{clustering-based} CF (CBCF) method only with ratings given by users, which is thus easy to implement. That is, we design such a simple clustering-based approach with no further prior information while improving the recommendation accuracy. To this end, in this paper,  we introduce the CBCF method using an incentivized/penalized user (IPU) model in improving the performance of recommender systems in terms of \textit{precision}, \textit{recall}, and $F_1$ score. More specifically, we present the CBCF method by carefully exploiting different preferences among users along with clustering. Our proposed method is built upon a predicted rating matrix-based clustering that
can drastically reduce the processing overhead of clustering.
In our CBCF method, we aim to select items to be recommended for users along with clustering. 
To this end, users are divided into several clusters based on the actual rating data
and Pearson correlation coefficient. Then, items are regarded as more important or less important depending on the clusters that the users belong to. Afterwards, we give each item an \textit{incentive}/\textit{penalty} according to the preference tendency by users within the same cluster.
The main contributions of our work are summarized as follows.
\begin{itemize}
\item An easy-to-implement CBCF method using the IPU model is proposed to further enhance the performance related to UX.

\item To design our CBCF method, we first formulate a constrained optimization problem, in which we aim to maximize the \textit{recall} (or equivalently $F_1$ score) for a given \textit{precision}.

\item We numerically find the amount of incentive/penalty that is to be given to each item according to the preference tendency by users within the same cluster.

\item We evaluate the performance of the proposed method via extensive experiments and demonstrate that $F_1$ score of the CBCF method using the IPU model is improved compared with the baseline CF method without clustering, while \textit{recall} for given (fixed) \textit{precision} can be significantly improved by up to about 50\%.
\end{itemize}

The remainder of this paper is organized as follows.
Related work to our contributions is presented in Section~\ref{RelatedWorks}.
Some backgrounds are presented in Section~\ref{Background}. The overview of our proposed CBCF using the IPU model and the problem definition are described in Section~\ref{ProblemFormulation}. The implementation details of our CBCF method are shown in Section~\ref{ProposedAlgorithm}. The datasets are described in Section~\ref{SystemModel}, and the performance is analyzed via experiments in Section~\ref{SimulationResults}.
Finally, we summarize our paper with some concluding remarks in Section~\ref{Conclusions}.

\section{Related Work}
\label{RelatedWorks}

The method that we propose in this paper is related to four broader areas of research, namely CF approaches in recommender systems, various clustering methods, clustering-based recommender systems, and several studies on the recommender systems that analyzed the performance metrics such as \textit{precision} and \textit{recall}. 

{\bf CF-aided recommender systems.} CF is one of the most popular techniques used by recommender systems, but has some shortcomings vulnerable to data sparsity and cold-start problems~\cite{Guo14}. If the data sparsity problem occurs with insufficient information about the ratings of users on items, then the values of predicted preference become inaccurate. Moreover, new users or items cannot be easily embedded in the CF process based on the rating information. There have been a plenty of challenges tackling these two problems~\cite{Bobadilla12, Sobhanam13}. On the other hand, some of studies focused on how to improve prediction accuracy of CF-aided recommender systems~\cite{Cai14, Liu14, Huang15}. In~\cite{Liu14, Huang15}, new similarity models were presented by using proximity impact popularity and Jaccard similarity measures, respectively. In~\cite{Cai14}, a typicality-based CF method, termed TyCo, was shown by taking into account typicality degrees. Recently, serendipitous CF-aided recommender systems received an attention, where surprising and interesting items are recommended to users~\cite{Lu12, Oku11, Adamopoulos11}.

{\bf Clustering methods.} Clustering has been widely used in diverse data mining applications: clustering algorithms such as $k$-Means and density-based spatial clustering of applications with noise (DBSCAN) were implemented in~\cite{c17} to monitor game stickiness; a novel objective function based on the entropy was proposed in~\cite{c18} to cluster different types of images; a cluster validity index based on a one-class classification method was presented in~\cite{c19} by calculating a boundary radius of each cluster using kernel functions; a modified version of mean shift clustering for one-dimensional data was proposed in~\cite{c20} to meet the real-time requirements in parallel processing systems; and a new criterion, called the cluster similar coefficient (CSC), was introduced in~\cite{c21} to determine the suitable number of clusters, to analyze the non-fuzzy and fuzzy clusters, and to build clusters with a given CSC.

{\bf Clustering-based recommender systems.} There has been diverse research to enhance recommendation accuracy
by means of clustering methods~\cite{Huang14, Yin14, Guerraoui15, Koohi16}.
In~\cite{Huang14}, CF and content-based filtering methods were conducted by finding
similar users and items, respectively, via clustering, and then personalized recommendation to the target user was made. As a result, improved performance on the \textit{precision}, \textit{recall}, and $F_1$ score was shown.
Similarly as in~\cite{Huang14}, communities (or groups) were discovered in~\cite{Yin14} before the application of matrix factorization to each community. 
In~\cite{Guerraoui15}, social activeness and dynamic interest features were exploited to find similar communities by item grouping, where items are clustered
into several groups using cosine similarity. As a result of grouping,
the $K$ most similar users based on the similarity measure were selected for recommendation.
The performance of user-based CF with several clustering
algorithms including $K$-Means, self-organizing maps (SOM), and \textit{fuzzy $C$-Means (FCM)} clustering methods was shown in~\cite{Koohi16}. It was shown that
user-based CF based on the FCM has the best performance
in comparison with $K$-Means and SOM clustering methods. Moreover, several clustering approaches were studied in CF-aided recommender systems: heterogeneous evolutionary clustering was presented in~\cite{c26} by dividing individuals with similar state values into the same cluster according to stable states; another dynamic evolutionary clustering was shown in~\cite{c27} by computing user attribute distances; and more recently, dynamic evolutionary clustering based on time weight and latent attributes was proposed in~\cite{c28}.

\textbf{Performance analysis in terms of \textit{precision} and \textit{recall}}. Performance metrics related to UX such as \textit{precision}, \textit{recall}, and $F_1$ score have been widely adopted for evaluating the accuracy of recommender systems~\cite{Wu16, Jiaa15, YLiu15, Wu15}.
In~\cite{Jiaa15}, time domain was exploited in designing CF algorithms by analyzing the inter-event time distribution of human behaviors when similarities between users or items are calculated.
In addition, performance on the accuracy of other various recommender systems was analyzed in~\cite{Wu16, YLiu15, Wu15} with respect to \textit{precision} and \textit{recall}.

\section{Backgrounds}
\label{Background}
In this section, we summarize both preference prediction based on several CF algorithms and two clustering algorithms.
\subsection{Preference Prediction Methods}


Preference prediction methods using CF are divided into
memory-based and model-based approaches.
Memory-based approaches directly utilize volumes of historical data to predict a rating
on a target item and provide recommendations for active users.
Whenever a recommendation task is performed, the memory-based approaches need to load all the data into
the memory and implement specific
algorithms on the data. On the other hand, model-based approaches leverage certain data mining methods to
establish a prediction model based on the known data. Once a model is obtained, it does
not need the raw data any more in the recommendation process~\cite{Yang16}. 

In our work, we adopt memory-based approaches for our CBCF method. Although model-based approaches offer the benefits of prediction speed and scalability, they have some practical challenges such as inflexibility and quality of predictions. More specifically, building a model is often a time- and resource-consuming process; and the quality of predictions depends heavily on the way that a model is built.

\subsubsection{User/Item-Based CF}

There are two major memory-based CF algorithms, i.e., $user$-based and $item$-based
algorithms.
In user/item-based CF, we make a prediction for an active user, $u$, on
a certain item $i$ after finding similar users/items, respectively.
Generally, in $user$-based CF, a correlation-based similarity is used for computing a user similarity
and then a weighted sum of other users' ratings are used for making a prediction.
In \textit{item}-based CF, a cosine-based similarity and a simple weighted average
can also be used for computing an item similarity and making a prediction, respectively.
For more detailed process of both CF algorithms, we refer to~\cite{Su09}.


\subsection{Clustering}

Among various clustering methods such as SOM, K-Means, FCM, and spectral
clusterings, we select spectral clustering and FCM, which have been widely known to ensure satisfactory performance. We briefly explain these two algorithms as follows.

\textit{Spectral} clustering is based on the spectrum of an affinity matrix.
In the affinity matrix, an affinity value between two objects (i.e., items) increases or decreases
when the similarity between two objects is high or small, respectively.
The Gaussian similarity function for quantifying the similarity between two objects
is widely used to construct the affinity matrix.\footnote{The Gaussian similarity function
	is given by $s(x_i,x_j)=e^{\frac{-{\lVert
x_i-x_j\rVert}^2}{2\sigma^2}}$, where $\sigma$ controls the width of the
	neighborhoods~\cite{Luxburg07}.}
After obtaining the affinity matrix, we find the corresponding eigenvectors/eigenvalues to
group objects into several clusters.
Finally, $spectral$ clustering divides objects based on
the eigenvectors/eigenvalues.
There are various strategies for object division (refer to~\cite{Luxburg07} for the details).
While $spectral$ clustering is simple to implement by
a standard linear algebra software tool, it is known to significantly outperform
traditional clustering algorithms such as $K$-Means clustering~\cite{Luxburg07}.

FCM clustering~\cite{fuzzycmeans} allows each object to be the member of all clusters with different degrees of fuzzy membership by employing a coefficient $w_{ij}^m$ that links an object $x_i$ to a cluster $c_j$, where $m$ is the hyper-parameter that controls how fuzzy the cluster will be. The higher $m$ is, the fuzzier the cluster will be. FCM clustering first initializes coefficients of each point at random given a number of clusters. Then, the following two steps are repeated until the coefficients' change between two iterations is less than a given sensitivity threshold: 1) Computing the centroid for each cluster and 2) Recomputing coefficients of being in the clusters for each point.


\section{Problem Formulation}
\label{ProblemFormulation}
In this section, we define and formulate our problem with a motivating example.
\subsection{Problem Definition}

The contribution of our work is to make a proper decision with which items
should be recommended or not under the same MAE or RMSE in terms of improving UX  (i.e., \textit{recall} (or equivalently $F_1$ score) for a given \textit{precision}).
For example, suppose that there are two items with the same
predicted preference value given by $3.9$.
If a recommender system only suggests items whose predicted preference
is over 4.0, then above two items will be dropped by the system.
However, there may be some users who are satisfied with the items, and thus
UX will decrease in this case.
In order to enhance the UX, we give each item an \textit{incentive} or \textit{penalty}
according to the preference tendency by users.
To this end, we $cluster$ users into some groups and make a decision on
which items are given the \textit{incentive/penalty} based on a
group that users belong to.

\begin{figure}[t]
\centering
\includegraphics[width=3.3in]{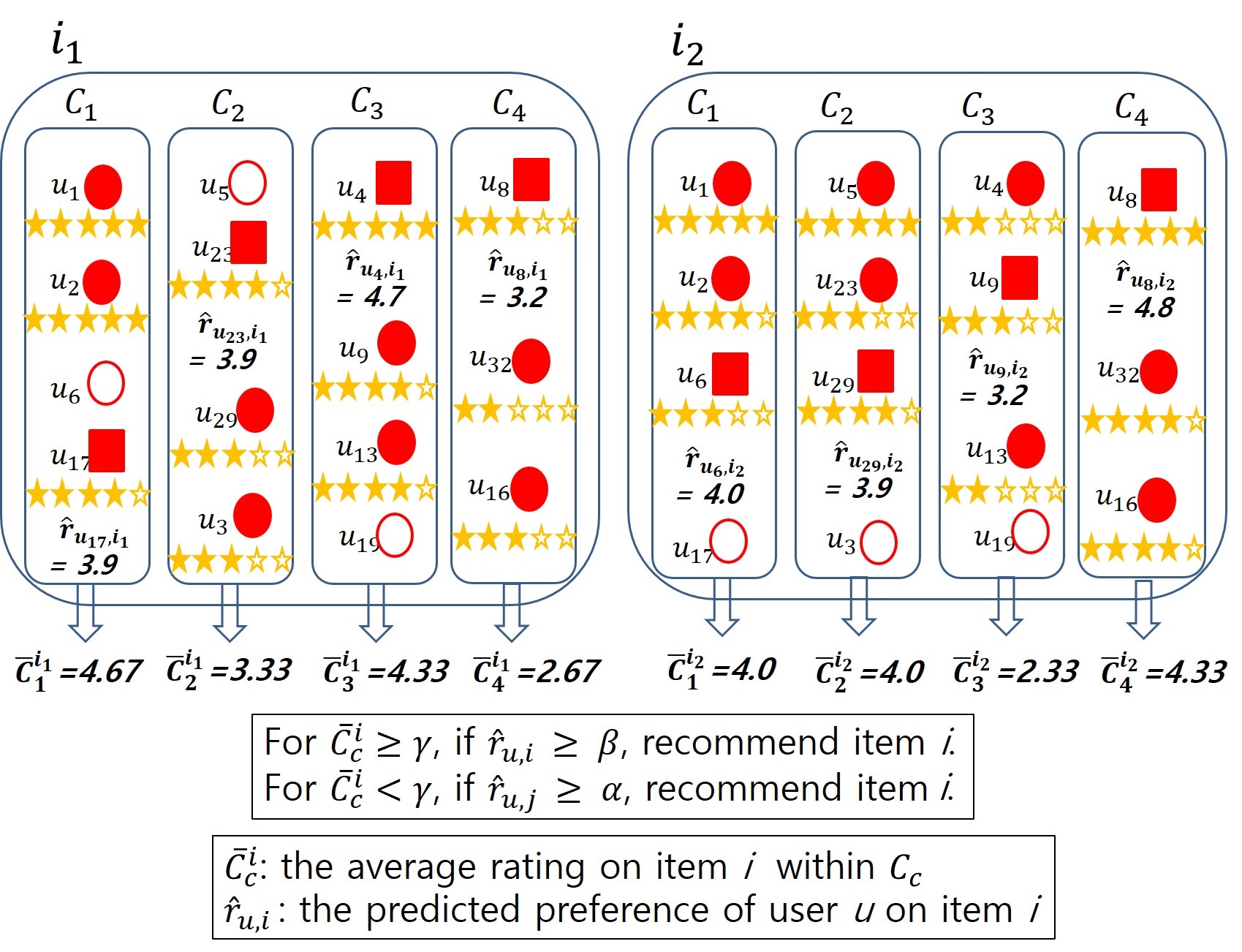}
\caption{An example of the proposed CBCF method with the IPU model, where two items and four clusters are assumed. Here, colored square items and colored circular items represent test data and training data, respectively.}
\label{example}
\end{figure}

Fig.~\ref{example} shows an example of our proposed CBCF method
with the IPU model, where two items and four clusters are assumed.
Users are assumed to be grouped into four clusters, i.e.,
$C_1$, $C_2$, $C_3$, and $C_4$.
From the figure, it can be seen that four users $u_1$, $u_2$, $u_6$, and $u_{17}$
belong to cluster $C_1$.
Here, colored square items and colored circular items represent test data and training
data, respectively.
We first denote $\hat{r}_{u,i}$ and $r_{u,i}$ as the $predicted$ preference 
and $real$ preference, respectively, of user $u$ on item $i$, where memory-based and model-based CF approaches can be employed for rating prediction (refer to Section~\ref{ProposedAlgorithm} for more details).
Then as illustrated in Fig.~\ref{example}, we have the $real$ preference $r_{u_{17},i_{1}} = 4.0$
and its $predicted$ preference $\hat{r}_{u_{17},i_{1}} = 3.9$.
Items that each user $u$ already rated along with the real preference are colored with red, whereas the others
are not.
For example, in cluster $C_1$, $i_1$ was rated as $5.0$, $5.0$, and
$4.0$ stars by
users $u_1$, $u_2$, and $u_{17}$, respectively, thus resulting in $r_{u_1,i_1}=5.0$,
$r_{u_2,i_1}=5.0$, and $r_{u_{17},i_{1}}=4.0$.
In the same cluster, users $u_1$, $u_2$, and $u_6$ rated $i_{2}$ as 5, 4, and 3 stars,
respectively, resulting in $r_{u_1,i_{2}}=5.0$, $r_{u_2,i_{2}}=4.0$, and
$r_{u_6,i_{2}}=3.0$.
Let us now denote $\bar{C}_c^i$ as the average preference on item $i$ of users within cluster $C_c$.
More specifically, $\bar{C}_{c}^{i}$ can be expressed as
\begin{eqnarray}
\bar{C}_{c}^{i} = \frac{\sum_{u \in U_{i,c}} r_{u,i}}{\lvert U_{i,c} \rvert},
\end{eqnarray}
where $U_{i,c}$ is the set of users who rated item $i$ within cluster $C_c$ and
$\lvert \cdot \rvert$ is the cardinality of a set.
Then, as shown in Fig.~\ref{example},
the average preference $\bar{C}_1^{i_1}$ of item $i_1$ rated by users within $C_1$ is given by $4.67$.
Similarly, $\bar{C}_2^{i_1}$ is given by $3.33$.

Based on the values of $\bar{C}_{c}^{i}$ in each cluster, we decide which
items should be recommended or not for user $u$ according to the following recommendation
strategy using the IPU model.
When the value of $\bar{C}_{c}^{i}$ is sufficiently large, i.e., $\bar{C}_{c}^{i} \geq \gamma$,
item $i$ is given an incentive, where $\gamma > 0$ indicates a system parameter
that is to be optimized later.
Otherwise (i.e., if  $\bar{C}_{c}^{i} < \gamma$), item $i$ having small
$\bar{C}_{c}^{i}$ gets a penalty.
System parameters $\alpha$ and $\beta$ are used as thresholds for giving a penalty and an incentive, respectively, in our method and are set to certain positive values,
where $\alpha \geq \beta$.
For example, suppose that $\alpha = 4.5$, $\beta=3.5$, and $\gamma = 3.0$.
Then, in Fig.~\ref{example}, $i_{1}$ will be recommended to $u_{19}$ but $i_2$ will not be recommended to $u_{19}$ if the predicted preferences of $i_{1}$ and $i_{2}$ (i.e., $\hat{r}_{u_{19},i_{1}}$ and
$\hat{r}_{u_{19},i_{2}}$) are $3.8$ and $4.2$, respectively.
This is because $\bar{C}_{3}^{i_1}$ $(=4.33)$ is larger than $\gamma$ $(=3.0)$
and $\hat{r}_{u_{19},i_{1}}$ $(=3.8)$ is also larger than $\beta$ $(=3.5)$.
In the case of $i_2$, however, $u_{19}$ does not receive recommendation since
$\bar{C}_{3}^{i_2}$ $(=2.33)$ is smaller than $\gamma$ as well as $\hat{r}_{u_{19},i_{2}} <
\alpha$.
In short, 
a decision on recommendation can be changed depending on the preference tendency of each
user obtained from clustering.

Algorithm~\ref{psuedoExample} describes our CBCF method using the IPU model.
From Algorithm~\ref{psuedoExample}, it is observed that items
rated over $\beta$ are just recommended when $\bar{C}_{c}^{i} \geq \gamma$.
If $\bar{C}_{c}^{i} < \gamma$, then only items whose predicted preference
is larger than $\alpha$ are recommended.

\begin{algorithm}[t!]                    
\caption{Proposed CBCF using the IPU model}          
\label{psuedoExample}
\SetLine
\eIf{$\bar{C}_{c}^{i} \geq \gamma$}
{
	\uIf{$\hat{r}_{u,i} \geq \beta$}
	{
		Recommend item $i$ to user $u$\;
	}
	\lElse
	{
		Drop item $i$\;
	}
}
{
	\uIf{$\hat{r}_{u,i} \geq \alpha$}
	{
		Recommend item $i$ to user $u$\;
	}
	\lElse
	{
		Drop item $i$\;
	}
}
\end{algorithm}

As mentioned before, we use the $precision$, $recall$, and $F_1$ score for performance
evaluation.
These three performance metrics can be expressed as functions of true positive $(tp)$,
true negative $(tn)$, false positive $(fp)$, and false negative $(fn)$.
Assume that we predict a $condition$ as true.
If the $condition$ is actually true (or false), then it is $tp$ (or $fp$).
If a $condition$ is predicted as false and the $condition$ is actually true (or false),
then it is $fn$ (or $tn$).

For given user $u$ and item $i$, the terms $tp$, $tn$, $fp$, and $fn$ are dependent on $\alpha$, $\beta$, and $\gamma$,
and thus are given by
\begin{equation}
\begin{aligned}
\label{Eq:tp}
f_{tp}^{u,i}(\alpha, \beta, \gamma, \delta_{\mathrm{pref}}) \hspace{5cm}\\
= I_{[\gamma,\infty)}(\bar{C}_c^{u,i}) \cdot 
I_{[\beta,\infty)}(\hat{r}_{u,i}) \cdot 
I_{[\delta_{\mathrm{pref}},\infty)}(r_{u,i})	\\
+ 
I_{(0,\gamma)}(\bar{C}_c^{u,i}) \cdot
I_{[\alpha,\infty)}(\hat{r}_{u,i}) \cdot
I_{[\delta_{\mathrm{pref}},\infty)}(r_{u,i}),	\\	
f_{fp}^{u,i}(\alpha, \beta, \gamma, \delta_{\mathrm{pref}}) \hspace{5cm}\\
= I_{[\gamma,\infty)}(\bar{C}_c^{u,i}) \cdot
I_{[\beta,\infty)}(\hat{r}_{u,i}) \cdot
I_{(0,\delta_{\mathrm{pref}})}(r_{u,i})	\\
+  
I_{(0,\gamma)}(\bar{C}_c^{u,i}) \cdot
I_{[\alpha,\infty)}(\hat{r}_{u,i}) \cdot
I_{(0,\delta_{\mathrm{pref}})}(r_{u,i}),			\\	
f_{fn}^{u,i}(\alpha, \beta, \gamma, \delta_{\mathrm{pref}}) \hspace{5cm}\\
= I_{[\gamma,\infty)}(\bar{C}_c^{u,i}) \cdot
I_{(0,\beta)}(\hat{r}_{u,i}) \cdot
I_{[\delta_{\mathrm{pref}},\infty)}(r_{u,i})	\\
+ 
				I_{(0,\gamma)}(\bar{C}_c^{u,i}) \cdot
				I_{(0,\alpha)}(\hat{r}_{u,i}) \cdot
				I_{[\delta_{\mathrm{pref}},\infty)}(r_{u,i}),		\\
		f_{tn}^{u,i}(\alpha, \beta, \gamma, \delta_{\mathrm{pref}}) \hspace{5cm}\\
 = I_{[\gamma,\infty)}(\bar{C}_c^{u,i}) \cdot
				I_{(0,\beta)}(\hat{r}_{u,i}) \cdot
				I_{(0,\delta_{\mathrm{pref}})}(r_{u,i})	\\
 + 
				I_{(0,\gamma)}(\bar{C}_c^{u,i}) \cdot
				I_{(0,\alpha)}(\hat{r}_{u,i}) \cdot
				I_{(0,\delta_{\mathrm{pref}})}(r_{u,i}),
\end{aligned}
\end{equation}
respectively, where $I_A(x)$ is the indicator function of set $A$ and $\delta_{\mathrm{pref}}$
is a threshold value for determining whether a user really
satisfies with the corresponding item.\footnote{Note that $\delta_{\mathrm{pref}}$
is generally set to $4.0$ (or $8.0$) in case of a five-point scale (or a ten-point scale).}
Then, it follows that $f_{tp}^{u,i}=1$ if $\bar{C}_c^{u,i} \geq \gamma$, $\hat{r}_{u,i}
\geq \beta$, and $r_{u,i} \geq \delta_{\mathrm{pref}}$; $f_{tp}^{u,i}=1$ if $\bar{C}_c^{u,i} < \gamma$,
	$\hat{r}_{u,i}
\geq \alpha$, and $r_{u,i} \geq \delta_{\mathrm{pref}}$; and $f_{tp}^{u,i}=0$ otherwise.
In a similar fashion, $f_{fp}^{u,i}=1$ if $\bar{C}_c^{u,i} \geq \gamma$, $\hat{r}_{u,i}
\geq \beta$, and $r_{u,i} < \delta_{\mathrm{pref}}$; $f_{fp}^{u,i}=1$ if $\bar{C}_c^{u,i} < \gamma$,
	$\hat{r}_{u,i}
\geq \alpha$, and $r_{u,i} < \delta_{\mathrm{pref}}$; and $f_{fp}^{u,i}=0$ otherwise.
Moreover, $f_{fn}^{u,i}=1$ if $\bar{C}_c^{u,i} \geq \gamma$, $\hat{r}_{u,i}
< \beta$, and $r_{u,i} \geq \delta_{\mathrm{pref}}$; $f_{fn}^{u,i}=1$ if $\bar{C}_c^{u,i} < \gamma$,
	$\hat{r}_{u,i}
< \alpha$, and $r_{u,i} \geq \delta_{\mathrm{pref}}$; and $f_{fn}^{u,i}=0$ otherwise.
Finally, $f_{tn}^{u,i}$ is also counted similarly as above, but it is not used for computing
the $precision$, $recall$, and $F_1$ score.

Based on~(\ref{Eq:tp}), the $precision$ and $recall$ are given by\footnote{To simplify notations, $precision(\alpha,\beta,\gamma,\delta_\text{pref})$ and $precision(\alpha,\beta,\gamma,\delta_\text{pref})$ will be written as $precision$ and $recall$, respectively, if dropping the arguments $\alpha,\beta,\gamma$, and $\delta_\text{pref}$ does not cause any confusion.}
\begin{align}
\begin{split}
&precision(\alpha, \beta, \gamma, \delta_{\mathrm{pref}})
= \frac{\sum_{(u,i) \in \text{T}}
	f_{tp}^{u,i}(\alpha, \beta, \gamma, \delta_{\mathrm{pref}})}
			{\sum_{(u,i) \in \text{T}} f_{tp}^{u,i}(\alpha, \beta, \gamma,
					\delta_{\mathrm{pref}}) +
			 \sum_{(u,i) \in \text{T}} f_{fp}^{u,i}(\alpha, \beta, \gamma,
					 \delta_{\mathrm{pref}})}\\
&recall(\alpha, \beta, \gamma, \delta_{\mathrm{pref}})
=  \frac{\sum_{(u,i) \in \text{T}}
	f_{tp}^{u,i}(\alpha, \beta, \gamma, \delta_{\mathrm{pref}})}
			{\sum_{(u,i) \in \text{T}} f_{tp}^{u,i}(\alpha, \beta, \gamma,
					\delta_{\mathrm{pref}}) +
			 \sum_{(u,i) \in \text{T}} f_{fn}^{u,i}(\alpha, \beta, \gamma,
					 \delta_{\mathrm{pref}})},
\end{split}
\end{align}
where $T$ represents the set of test data used for measuring $precision$ and $recall$.
Due to the fact that the $F_1$ score is the harmonic mean of $precision$ and $recall$,
it is defined as
\begin{eqnarray}
F_1(\alpha, \beta, \gamma, \delta_{\mathrm{pref}}) = \frac{2precision \times recall}{precision + recall}.
\label{Eq:F1}
\end{eqnarray}

Let us recall the example in Fig.~\ref{example}, where $\alpha=4.5$, $\beta=3.5$,
and $\gamma=3.0$.
Square items representing the
test data are used for performance analysis.
Suppose that items rated over 4 stars are satisfactory for users, i.e.,
		$\delta_{\mathrm{pref}}
= 4.0$, which is a typical assumption in recommender systems~\cite{Xu13}.
Then, user $u_{17}$ should receive recommendation for item $i_1$, whereas user $u_8$ should not.
Users $u_{29}$ and $u_{8}$ are actually satisfied with item $i_2$.
Based on the test dataset in Fig.~\ref{example}, the terms
$tp$, $tn$, $fp$, and $fn$ are summarized in Table~\ref{tableTypeExample}.
For comparison, let us consider a baseline scenario where clustering is not exploited. To this end, we assume $\gamma=0$ and modify the recommendation
strategy so that item $i$ is recommended only if the predicted preference $\hat{r}_u^i$
is no less than $4.0$. In this case, the four terms \textit{tp}, \textit{fn}, \textit{fp}, and \textit{tn} are also depicted in
Table~\ref{tableTypeExample}.
Using the result of Table~\ref{tableTypeExample}, we are ready to compute the $precision$
and $recall$ for the two cases, i.e., $\gamma=0$ and $\gamma=3.0$, as follows.
\begin{itemize}
\item $\gamma=0$ (baseline): From Table~\ref{tableTypeExample},
it follows that $tp=2$, $fp=1$, and $fn=3$. Thus, using~(\ref{Eq:tp}),
we have $precision=2/3$ and $recall=2/5$.
\item $\gamma=3.0$ (proposed): Suppose that $\alpha=4.5$ and $\beta=3.5$.
From Table~\ref{tableTypeExample} and~(\ref{Eq:tp}), it follows that
$tp=4$, $fp=1$, and $fn=1$. Hence, we have $precision=4/5$ and
$recall=4/5$.
\end{itemize}
Consequently, performance on the $precision$ and $recall$ can be improved
by properly adjusting the system parameters $\alpha$, $\beta$, and $\gamma$
under our IPU model when items are grouped into multiple clusters.


\subsection{Formulation}

\begin{table}[!t]
\renewcommand{\arraystretch}{1.3}
\caption{An example of \textit{tp}, \textit{fn}, \textit{fp}, and \textit{tn} when $\gamma=0$ and $\gamma=3$.}
\label{tableTypeExample}
\centering
\begin{tabular}{l | l | l | l}
\multirow{4}{1.2cm}
{$\gamma=0$ (baseline)}	& 						& item $i_1$							& item $i_2$	\\
\hline
					& Recommended items		& $u_{4}$ $\Rightarrow$ $tp$	& $u_{6}$ $\Rightarrow$ $fp$	\\
					& 						& 								& $u_{8}$ $\Rightarrow$ $tp$	\\
\cline{2-4}
					& Non-recommended items	& $u_{17}$ $\Rightarrow$ $fn$	& $u_{9}$ $\Rightarrow$ $tn$ \\
					& 						& $u_{23}$ $\Rightarrow$ $fn$	& $u_{29}$ $\Rightarrow$ $fn$	\\
					& 						& $u_{8}$ $\Rightarrow$ $tn$	&	\\
\hline
\hline
\multirow{4}{1.2cm}
{$\gamma=3.0$ (proposed)}	& \multirow{2}{*}{Recommended items}		& $u_{17}$ $\Rightarrow$ $tp$	& $u_{6}$ $\Rightarrow$ $fp$	\\
					& 						& $u_{4}$ $\Rightarrow$ $tp$	& $u_{29}$ $\Rightarrow$ $tp$	\\
					& 						& 								& $u_{8}$ $\Rightarrow$ $tp$	\\
\cline{2-4}
					& Non-recommended items	& $u_{23}$ $\Rightarrow$ $fn$	& $u_{9}$ $\Rightarrow$ $tn$ \\
					& 						& $u_{8}$ $\Rightarrow$ $tn$	& \\
\hline
\end{tabular}
\end{table}

It is worth noting that the
$precision$, $recall$, and $F_1$ score vary significantly according to the change of
$\alpha$, $\beta$, and $\gamma$.
For this reason, we aim at finding the optimal $\alpha$, $\beta$, and $\gamma$
such that the $F_1$ score (or \textit{recall}) is maximized.
We thus formulate a new constrained optimization problem as follows:\footnote{Since the
	parameter $\delta_{\mathrm{pref}}$
is generally set to a certain value, $\delta_{\mathrm{pref}}$
will be dropped from the argument of each function to simplify notations if dropping it does not cause any confusion.}
\begin{eqnarray}
\begin{aligned}
& \underset{\alpha, \beta, \gamma}{ \text{maximize}}
& & F_1(\alpha, \beta, \gamma) \,\, \mathrm{or } \,\,recall(\alpha, \beta, \gamma)\\
& \text{subject to}
& & precision(\alpha, \beta, \gamma) \geq \delta_{\mathrm{precision}} \\
& & & \alpha \geq \beta,
\end{aligned}
\label{Eq:formula}
\end{eqnarray}
where
$\delta_{\mathrm{precision}}$ is a pre-defined threshold value for $precision$
and is set to a certain value appropriately according to various types
of recommender systems.
Equation~(\ref{Eq:formula}) can be also easily modified for different purposes.
For example, we can find the optimal $\alpha$, $\beta$, and $\gamma$ such that
$precision(\alpha, \beta, \gamma)$ is maximized under $recall(\alpha, \beta, \gamma) \geq
\delta_{\mathrm{recall}}$
or $recall(\alpha, \beta, \gamma)$ is maximized under $precision(\alpha, \beta, \gamma) \geq
\delta_{\mathrm{precision}}$, where $\delta_{\mathrm{recall}}$ is a pre-defined
threshold value for $recall$.
Hence, the $precision$, $recall$, and $F_1$ score can be improved by not only clustering items
but also optimally finding parameters $\alpha$, $\beta$, and $\gamma$ in our CBCF method using the IPU model.

\section{Proposed Method}
\label{ProposedAlgorithm}

The CBCF method recommends desirable items
according to the result of item clustering and the preference tendency of each
user using our IPU model.

The main contribution of our CBCF method using the IPU model is to
give either an incentive or a penalty to each item
based on $\bar{C}_c^i$ (the average preference on item $i$ of users within cluster $C_c$), which depends on the result
of clustering.
As mentioned before, since there are empty elements in the rating matrix $\mathbf{R}_{CBCF}$
that users have not rated or acccessed yet, the Euclidian distance between user vectors (i.e., row vectors in $\mathbf{R}_{CBCF}$) cannot be accurately calculated.
Hence, we use the Pearson correlation coefficient (PCC) in our work.
PCC computes the correlation between two users' common ratings to measure
their similarity, and thus needs two common ratings at least.
PCC between two users, $u_1$ and $u_2$, is calculated as
\begin{flalign}
s({u_1},{u_2})=\frac{\sum_{i \in I_{u_1} \cap I_{u_2}} (r_{{u_1},i}-\bar r_{u_1})\cdot(r_{{u_2},i}-\bar
		r_{u_2})}{\sqrt{\sum_{i \in I_{u_1} \cap I_{u_2}} (r_{{u_1},i}-\bar r_{u_1})^2} \cdot \sqrt{\sum_{i
	\in I_{u_1} \cap I_{u_2}} (r_{{u_2},i}-\bar r_{u_2})^2}},
\end{flalign}
where $I_{u_1}$ and $I_{u_2}$ are the item sets rated by ${u_1}$ and ${u_2}$, respectively,
and $\bar r_{u_1}$ and $\bar r_{u_2}$ are the mean values of their ratings over the item set $I_{u_1} \cap
I_{u_2}$ that two users have commonly rated, respectively.
Here, $s({u_1},{u_2})$ ranges from $-1$ to $1$.
A correlation coefficient  close to $-1$ indicates a negative linear relationship,
and $s({u_1},{u_2})$ of 1 indicates a perfect positive linear relationship.

Let us turn our attention to the description of our
CBCF method in Algorithm~\ref{CBR2}.
First, the set of clusters, $C$, is obtained by the result of clustering where $c$ groups are generated,
and an $n \times m$ rating matrix $\mathbf{R}_{CBCF}$ is initialized (refer to lines 1--2 in Algorithm~\ref{CBR2}).
In the next step, we use a preference prediction method based on memory-based approaches along with $\mathbf{R}_{CBCF}$
and the resulting output is stored in $\hat{R}$ (refer to line 3).
More specifically, user/item-based CF algorithms are used to evaluate
the performance of our proposed CBCF method.
The threshold values $\alpha$, $\beta$, and $\gamma$ can be determined
by solving the optimization problem in~(\ref{Eq:formula}) via exhaustive search.
In the $for$ loop, the set $I_u$ is the items  of missing ratings in the test set for each user $u$ and the predicted ratings
in $I_u$ are assigned to $\hat{r}_{u,I_u}$, where $|I_u|$ denotes the cardinality of the set $I_u$.
Now, we decide which items are recommended or dropped for given $\alpha$, $\beta$,
and $\gamma$.
When $\hat{r}_{u,i} \geq \alpha$, the item $i$ is recommended to user $u$
regardless of the value of $\gamma$ as mentioned in Algorithm~\ref{psuedoExample} (refer to lines
11--12 in Algorithm~\ref{CBR2}).
However, when $\hat{r}_{u,i} < \alpha$, we have to check the value of threshold $\gamma$, which is to be compared with the average preference on a certain item of users in a cluster, denoted by $\bar{C}_{tmp}^{i}$.
When $\bar{C}_{tmp}^i <\gamma$, the item $i$
will not be recommended even if $\beta\le \hat{r}_{u,i}<\alpha$.
This is because we give a penalty to the item $i$ for $\bar{C}_{tmp}^{i} < \gamma$.
On the other hand, when $\hat{r}_{u,i}>\beta$ and $\bar{C}_{tmp}^{i} \geq \gamma$, the item $i$ will 
be recommended to user $u$ (refer to lines 13--14).
The item $i$ will be always dropped when $\hat{r}_{u,i} < \beta$ (refer to line 15).

\begin{algorithm}[t!]                      
\caption{CBCF using the IPU model}          
\label{CBR2}                           
\SetLine
Clusters $C \in\{C_1,\cdots,C_c\}$\;
Initialize the $n \times m$ rating matrix $\mathbf{R}_{CBCF}$\;
$\hat{R} \leftarrow$ a function of rating prediction with $\mathbf{R}_{CBCF}$\;
Initialize the threshold values $\alpha$, $\beta$, and $\gamma$\;
\For{$u\leftarrow 1$ \KwTo $n$}
{
	$I_u \leftarrow$ items  of missing ratings in the test set for user $u$\;
	$\hat{r}_{u, I_u} \leftarrow$ predicted rating values of $I_u$\;

	\For{$i\leftarrow 1$ \KwTo $|I_u|$}
	{
		$C_{tmp} \leftarrow$ a cluster to which user $u$ belongs\;
		$\bar{C}_{tmp}^{i}  \leftarrow$ average rating on item $i$ in $C_{tmp}$\;

		\uIf{$\hat{r}_{u,i} \geq \alpha$}
		{
			Recommend item $i$ to user $u$\;
		}
		\uElseIf {$\hat{r}_{u,i} \geq \beta$ \&\&
			$\bar{C}_{tmp}^{i} \geq \gamma$}
		{
			Recommend item $i$ to user $u$\;
		}
		\lElse {Drop item $i$\;}
	}
}
\end{algorithm}

Finally, we find $\alpha$, $\beta$, and $\gamma$ fulfilling~(\ref{Eq:formula}).
Algorithm~\ref{CBR2} is performed iteratively while varying the values of $\alpha$,
$\beta$, and $\gamma$. That is, lines 4--17 in Algorithm~\ref{CBR2} are iteratively
executed by  numerically optimizing  $\alpha$, $\beta$, and $\gamma$ according to~(\ref{Eq:formula}).

The CBCF method using the IPU model is summarized as follows:
\begin{itemize}
\item Suppose that the CBCF method decides whether a certain item $(i)$ is recommended to an active user $(u)$ or not under the IPC model based on clustering.
\item If the predicted preference is sufficiently large (i.e., $\hat{r}_{u,i} \geq \alpha$), then the
item $i$ is recommended to the user $u$.
\item If the predicted preference is not sufficiently large but the two conditions, i.e., $\hat{r}_{u,i} \geq \beta$
and $\bar{C}_{c}^{i} \geq \gamma$, are met, then the item $i$ is recommended to the user $u$,
where $\bar{C}_{c}^{i}$ is the average preference on item $i$ of users within cluster $C_c$.
\end{itemize}

\section{Dataset and Database Structure}
\label{SystemModel}

In this section, we describe our dataset and database (DB) structure. CBCF is utilized for non-cold-start users, but it will be empirically shown in Section~\ref{SimulationResults} how it is robust to more difficult situations including cold-start users.\footnote{In this paper, a cold-start user
is defined as the user who does not have enough rating information.
More than 20 ratings for each user are usually known as enough information~\cite{Pereira15}}.
We use the MovieLens 100K dataset\footnote{http://grouplens.org/datasets/movielens/.} with the following attributes:
\begin{itemize}
\item 100K dataset have 100,000 anonymous ratings
\item Ratings are made on a 5-star scale
\item There are 943 users in 100K dataset
\item There are 1,682 movies in 100K dataset
\item Each user has at least 20 ratings.
\end{itemize}
Note that the sparsity (i.e., the ratio of the number of missing cells in a rating matrix to the total number of cells) of the rating matrix obtained from the MovieLens 100K dataset is 93.7\%, which is high and often causes performance degradation. One popular solution to the data sparsity problem is the use of data imputation \cite{c35,c36,c37}, which includes the zero injection method~\cite{c35} in which zeros are given to some missing cells in a rating matrix and two matrix factorization-based methods ~\cite{c36,c37} that assign zeros or twos to all missing cells in a rating matrix. Even if such data imputation techniques are known to significantly improve the prediction accuracy, we do not employ them in our experiments since solving the data sparsity problem is not our primary focus. The DB structure for CBCF is described as follows.

Assume that there are a set of users, $U$,
and a set of items, $I$, in a recommender system as follows:
\begin{eqnarray}
U & \triangleq & \{u_1,u_2,\cdots , u_n\}, \nonumber \\
I & \triangleq & \{i_1,i_2,\cdots , i_m\},
\label{Eq:V}
\end{eqnarray}
where $n$ and $m$ represent the number of users and the number
of items, respectively.
Then, in the CBCF process, the rating matrix $\mathbf{R}_{CBCF}$ is defined as
\begin{eqnarray}
\label{eqn_R}
\mathbf{R}_{CBCF} =
\left( \begin{array}{ccccc}
        r_{1,1}     & r_{1,2}   & r_{1,3}   & \ldots &  r_{1,m} \\

        r_{2,1}     & r_{2,2}   & r_{2,3}   & \ldots &  r_{2,m} \\
        
        r_{3,1}     & r_{3,2}   & r_{3,3}   & \ldots &  r_{3,m} \\
        
        \vdots  & \vdots    & \vdots    & \ddots    & \vdots    \\

        r_{n,1}     & r_{n,2}   & r_{n,3}   & \ldots &  r_{n,m} 
\end{array} \right),
\end{eqnarray}
where $r_{u,i}$ is the rating of user $u$ on item $i$ for $u \in \{1,\cdots,n\}$
and $i \in \{1,\cdots,m\}$.
Note that $\mathbf{R}_{CBCF}$ can be either the users' explicit ratings or the users' implicit
preferences.
If user $u$ has not rated or accessed item $i$ yet, then $r_{u,i}$ remains empty.

The user set $U$ is grouped into several
clusters and a \textit{user cluster} is a set of similar users in the rating matrix
$\mathbf{R}_{CBCF}$.
In order to cluster $U$, we define $n$ user vectors, each of which consists of $m$ elements, which are given by
\begin{eqnarray}
\label{Eq:vector}
\mathbf{U}_b = [r_{b,1}, r_{b,2}, \cdots , r_{b,m}]
\end{eqnarray}
for $b \in \{1,\cdots,n\}$.
Suppose that
$n$ user vectors are clustered into $c$ user groups,\footnote{For clustering, it is of importance how to determine the number of clusters. This is heavily dependent on
the characteristics  of recommender systems and thus is beyond the scope of this paper.} where the set of clusters, $C$, is denoted by
\begin{eqnarray}
C = \{C_1, C_2, \cdots , C_{c}\}.
\end{eqnarray}
In this case, one can say that the users within a cluster are relatively closer than
other users not in the same cluster from the
viewpoint of users' preferences.
For example, assume that there are four user vectors
given by $\mathbf{U}_1 = [2, 0, 1, 0]$,
$\mathbf{U}_2 = [0, 4, 0, 2]$,
$\mathbf{U}_3 = [3, 0, 2, 0]$,
and $\mathbf{U}_4 = [0, 3, 0, 2]$.
Let us divide the four vectors into two clusters.
Then, $\mathbf{U}_1$ and $\mathbf{U}_3$ will be grouped into one cluster and are considered as similar users by the users' ratings because
the Euclidian distance between $(\mathbf{U}_1, \mathbf{U}_3)$ is closer than
that made from other combinations including $(\mathbf{U}_1, \mathbf{U}_2)$,
$(\mathbf{U}_1, \mathbf{U}_4)$,
	$(\mathbf{U}_3, \mathbf{U}_2)$, and
	$(\mathbf{U}_3, \mathbf{U}_4)$.

The DB structure for CBCF is shown in
Table~\ref{table1}.
The DB consists of the following three fields: user ID,
item ID, and ratings.
For example, if item $i_1$ was enjoyed by user $u_1$ and was rated as 4.0, then
a new tuple `$u_1|i_1|4.0$' will be
inserted into the DB.

\begin{table}[!t]
\renewcommand{\arraystretch}{1.3}
\caption{DB structure of CBCF.}
\label{table1}
\centering
\begin{tabular}{c | c | c}
\hline
User ID		& Item ID	& Ratings $(\mathbf{R}_{CBCF})$	\\
\hline \hline
$u_1$		& $i_1$		& $r_{1,1}$		\\
\hline
$u_1$		& $i_2$		& $r_{1,2}$		\\
\hline
$u_1$		& $i_8$		& $r_{1,8}$		\\
\hline
$\vdots$	& $\vdots$	& $\vdots$		\\
\hline
$u_{n}$	&	$i_{m-4}$	& $r_{n,m-4}$	\\
\hline
$u_{n}$	&	$i_{m}$		& $r_{n,m}$	\\
\hline
\end{tabular}
\end{table}

\section{Experimental Results}
\label{SimulationResults}

In this section, we evaluate the performance of our proposed CBCF method using the IPU model in terms of
$precision$, $recall$, and $F_1$ score. In our experiments, unless otherwise stated, item-based CF is adopted in our proposed method since it shows better performance on the accuracy of recommendation for memory-based CF, which will be verified later in this section.
We use Apache Mahout\footnote{http://mahout.apache.org/.} whose goal is to build an environment for
performing downstream machine learning tasks such as CF, clustering, and classification.
It is assumed that the recommendation result is true when
the following conditions are met:
\begin{itemize}
\item The real rating of an item recommended to a user is 4.0 or 5.0.
\item The real rating of an item not recommended to a user is less than $4.0$.
\end{itemize}

\begin{figure}[t!]
\centering
\includegraphics[width=3.5in]{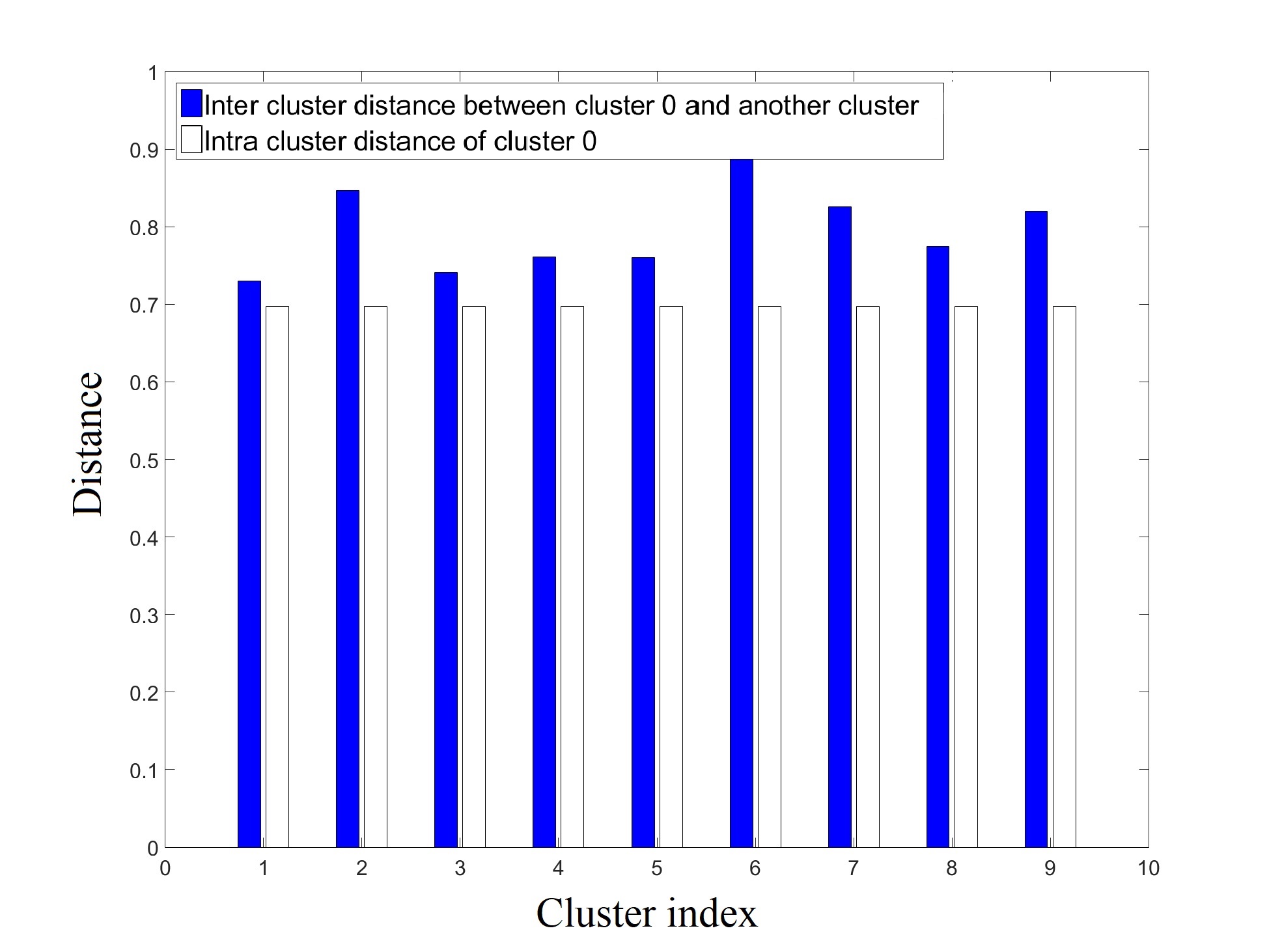}
\caption{Comparison of the inter-cluster and intra-cluster Euclidean distances. \label{clusterComp}}
\end{figure}

In our experiments, the number of clusters for both \textit{spectral} and FCM clustering algorithms is set to $c=10$; the fuzzy degree $m$ of FCM clustering is set to 2 according to~\cite{fuzzy}; and the convergence threshold of FCM clustering is set to $10^{-4}$. In the FCM clustering, an object is assigned to such a cluster that has the highest coefficient. In our subsequent experiments, we adopt \textit{spectral} clustering by default unless otherwise stated.
Fig.~\ref{clusterComp} compares the inter-cluster Euclidean distances with
the intra-cluster Euclidean distances in order to show the validity of clustering.
The values of PCC range between $-1.0$ and $1.0$, where $1.0$ and $-1.0$ imply that two objects (e.g., users) have
the highest positive and negative correlations, respectively.
Hence, since most clustering algorithms do not employ any negative correlation, the value of PCC between two users $u_1$ and $u_2$, namely $s(u_1,u_2)$, is shifted as follows:

\begin{eqnarray}
\begin{aligned}
&s(u_1,u_2) \leftarrow  1-s(u_1,u_2)  \text{ for } s(u_1,u_2) \in [0, 1] \\
&s(u_1,u_2) \leftarrow  -(s(u_1,u_2)-1)  \text{ for } s(u_1,u_2) \in [-1, 0). 
\end{aligned}
\end{eqnarray}

Then, a value close to 0 indicates a highly positive correlation while a value close to 2 corresponds to a highly negative correlation.
As shown in Fig.~\ref{clusterComp}, it is observed that
the intra-cluster distance is smaller than the inter-cluster distances from the perspective of cluster 0.
It thus reveals that our PCC-based clustering works appropriately.

\begin{figure}[t!]
\centering
\includegraphics[width=3.5in]{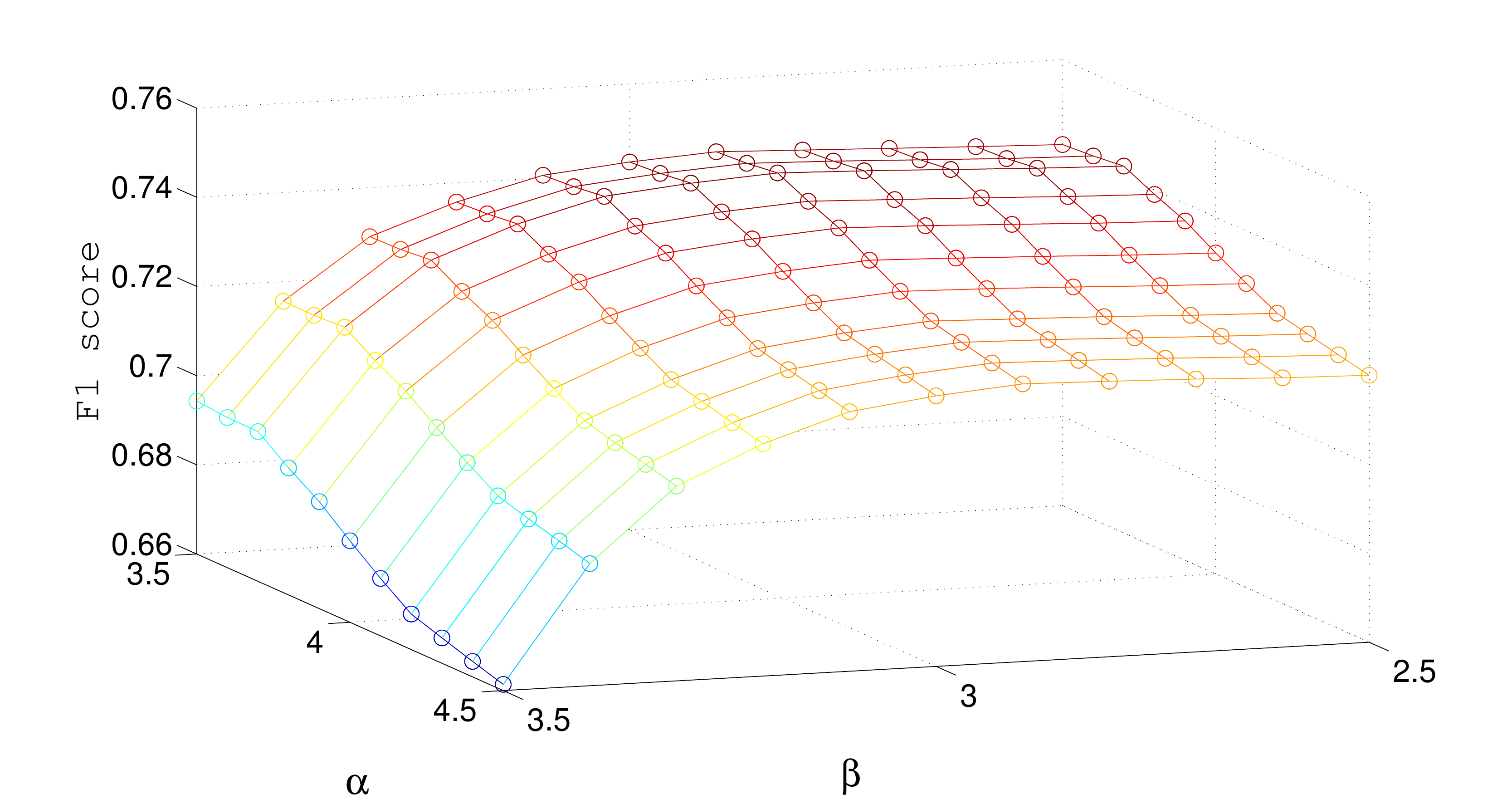}
\caption{$F_1$ score over $\alpha$ and $\beta$ when $\gamma = 3.4$. \label{F1ScoreItemCF}}
\end{figure}

Fig.~\ref{F1ScoreItemCF} shows the effect of $\alpha$ and $\beta$, which correspond to thresholds for giving a penalty and an incentive, respectively, on the $F_1$ score when
another threshold $\gamma$ is set to $3.4$. We note that the proposed CBCF method using the IPU model
has the maximum $F_1$ score (=0.7451) when $\alpha=3.7$ and $\beta=2.9$.
It is observed that the $F_1$ score decreases
as $\alpha$ and $\beta$ increase since the decreasing rate of $recall$
is larger than the increasing rate of $precision$ with increasing $\alpha$ and $\beta$.
More specifically, if both $\alpha$ and $\beta$ are large, then $precision$ and $recall$ tend to increase and decrease, respectively, since
fewer items are recommended.
However, due to the fact that the decrement of $recall$ is faster than
the increment of $precision$, the $F_1$ score gets reduced accordingly.
For example, in Fig.~\ref{F1ScoreItemCF}, it is seen that $precision = 0.6595$ and $recall = 0.8564$ when
$\alpha=3.7$, $\beta=2.9$, and $\gamma=3.4$,
while $precision = 0.6853$ and $recall = 0.076$ when
$\alpha=4.4$, $\beta=4.4$, and $\gamma=3.4$.

\begin{figure}[t!]
\centering
\includegraphics[width=3.5in]{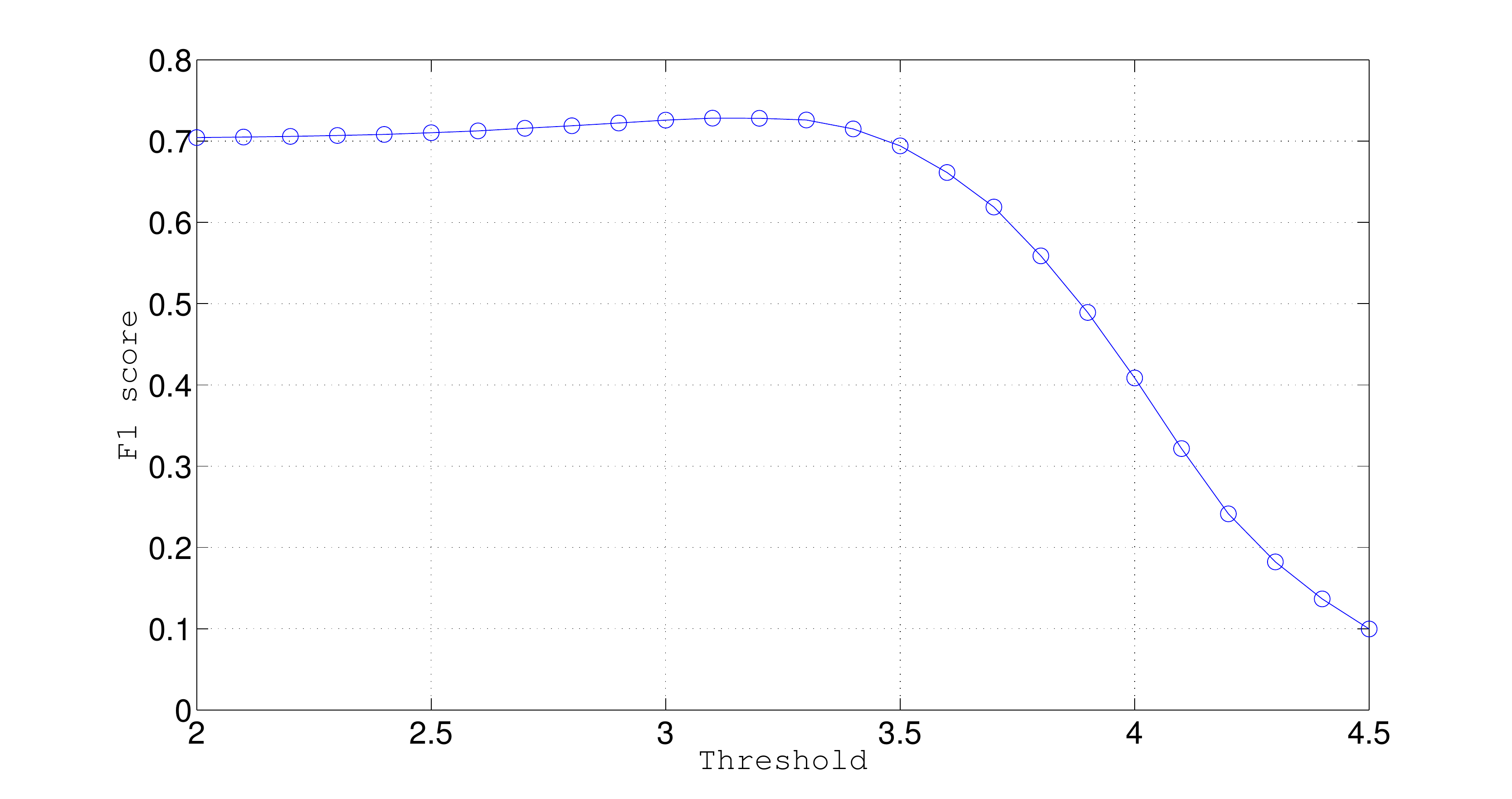}
\caption{$F_1$ score over the recommendation threshold when item-based CF is adopted.} 
\label{F1OrgItemCF}
\end{figure}

Fig.~\ref{F1OrgItemCF} shows the $F_1$ score over the recommendation threshold when the baseline item-based CF method without clustering (i.e., $\gamma=0$) is adopted. In this baseline approach, if the predicted rating of a certain item
is larger than the recommendation threshold, then the corresponding
item is recommended to a user. If the real rating is over 4.0, then the recommendation
is regarded as valid.
As shown in this figure, the maximum of $F_1$ score
is 0.7282 when the threshold value is given by $3.1$.
It is shown that the overall tendency is similar to that in Fig.~\ref{F1ScoreItemCF},
but the $F_1$ score of the proposed method is increased by about 3\% compared to this baseline approach employing item-based CF.

Table~\ref{Maximum} shows the $recall$ and $F_1$ score
for given $precision$ when the proposed CBCF using the PIU model and the baseline method without clustering are used.
In the baseline item-based CF method, when the recommendation threshold is set to 4.0, the value of \textit{precision} is 0.7449 and the corresponding maximum $recall$ is 0.2815.
On the other hand, in the proposed method,
when $\alpha=3.9$, $\beta=2.1$, and $\gamma=4.2$, the maximum value of \textit{recall} is 0.4343.
This improvement is nearly 50\%.
That is, the proposed method has a remarkably higher $recall$ value compared to the baseline
under the same $precision$ as depicted in Table~\ref{Maximum}.
From Figs.~\ref{F1ScoreItemCF} and \ref{F1OrgItemCF}, and Table~\ref{Maximum}, it is shown that
the proposed method can achieve a great improvement with respect to
$recall$ or $F_1$ score for given $precision$. 

\begin{table}[!t]
\renewcommand{\arraystretch}{1.3}
\caption{The maximum \textit{recall} and $F_1$ score for given \textit{precision}.}
\label{Maximum}
\centering
\begin{tabular}{|c|c|c|c|c|}
\hline
\multirow{2} * {\textit{precision}}  & \multicolumn{2}{c|}{Baseline method} & \multicolumn{2}{c|}{Proposed method} \\ \cline{2-5}
 & \textit{recall} & $F_1$ score & \textit{recall} & $F_1$ score \\ \hline
0.7449 & 0.2815 & 0.4085 & 0.4343 & 0.5487 \\ \hline
0.7201 & 0.4565 & 0.5588 & 0.5706 & 0.6367 \\ \hline
0.7074 & 0.5499 & 0.6188 & 0.6842 & 0.6956 \\ \hline
0.6519 & 0.7914 & 0.7149 & 0.825 & 0.7283 \\ \hline
0.6036 & 0.9177 & 0.7282 & 0.9402 & 0.7352 \\ \hline
\end{tabular}
\end{table}

Generally, a small recommendation threshold value leads a low $precision$ and high $recall$,
and vice versa.
However, as mentioned before, as the threshold value becomes very large, the $F_1$ score
is rapidly decreased because the decreasing rate of $recall$ is faster than
the increasing rate of $precision$.

Instead of item-based CF, user-based CF can also be employed in our proposed CBCF method. When the parameters $\alpha$, $\beta$, and $\gamma$ are optimally found via exhaustive search  in the sense of maximizing the $F_1$ score, we evaluate the performance of our proposed CBCF method using the IPU model based on item-based CF and user-based CF methods in Tables~\ref{withItemBasedCF} and \ref{withUserBasedCF}, respectively, where both \textit{spectral} and FCM clustering algorithms are employed for non-cold-start users.
Based on the results, the following observations are made: i) the proposed method based on item-based CF achieves better performance on the $F_1$ score than the case of user-based CF and ii) using the proposed method based on FCM clustering is slightly superior to the case of \textit{spectral} clustering.

Moreover, we evaluate the performance of the proposed and baseline methods for more difficult situations having cold-start users whose number of rated items is less than 20, where item-based CF and \textit{spectral} clustering are used. Due to the fact that the MovieLens 100K dataset does not contain records for cold-start users, we modify the experimental setup according to~\cite{coldsetting}. Specifically, we first select users who have rated between 20–30 items as the testing set, consisting of 290 users, and make the number of rated items of each selected user in the range between 3 and 20 via random masking. The remaining 653 users from the original dataset is used as the training set. The results in Table~\ref{withItemBasedCF_cold} follow similar trends to those for non-cold-start users while the CBCF method provides gains over the baseline without clustering, where the three threshold values are optimally found in the sense of maximizing the $F_1$ score.


\begin{table}[!t]
\renewcommand{\arraystretch}{0.6}
\caption{Performance of the proposed method based on item-based CF, where both \textit{spectral} and FCM clustering algorithms are employed.}
\label{withItemBasedCF}
\centering
\begin{tabular}{|c|c|c|c|c|c|c|}
\hline
Clustering & $\gamma$ & $\alpha$ & $\beta$ & \textit{precision} & \textit{recall} & $F_1$ score \\ \hline
Spectral & 3.4	 &  3.7 & 2.9 & 0.6595	 & 0.8564 & 0.7451 \\ \hline
FCM & 3.5	 &  3.3 & 2.5 & 0.6625	 & 0.8639 & {\bf 0.7499} \\ \hline
\end{tabular}
\end{table}

\begin{table}[!t]
\renewcommand{\arraystretch}{0.6}
\caption{Performance of the proposed method based on user-based CF, where both \textit{spectral} and FCM clustering algorithms are employed.}
\label{withUserBasedCF}
\centering
\begin{tabular}{|c|c|c|c|c|c|c|}
\hline
Clustering & $\gamma$ & $\alpha$ & $\beta$ & \textit{precision} & \textit{recall} & $F_1$ score \\ \hline
Spectral & 3.5	 &  3.1 & 2.7 & 0.6309  & 0.8893 & 0.7382 \\ \hline
FCM & 3.3	 &  3.7 & 2.9 & 0.6448	 & 0.8730 &{\bf 0.7418} \\ \hline
\end{tabular}
\end{table}

\begin{table}[!t]
\renewcommand{\arraystretch}{0.6}
\caption{Performance of the proposed and baseline methods for cold-start users.}
\label{withItemBasedCF_cold}
\centering
\begin{tabular}{|c|c|c|c|}
\hline
Method &  \textit{precision} & \textit{recall} & $F_1$ score \\ \hline
Baseline method &  0.7085	 & 0.3552 & 0.4732 \\ \hline
Proposed method & 0.6793	 & 0.6934 & {\bf 0.6863} \\ \hline
\end{tabular}
\end{table}




\section{Concluding Remarks}\label{Conclusions}
In this paper, we proposed a CBCF method using the IPU model in recommender systems by carefully exploiting different preferences among users along with clustering. 
Specifically, in the proposed CBCF method, we formulated a constrained optimization problem in terms of maximizing the \textit{recall} (or equivalently $F_1$ score) for a given \textit{precision}. To this end, clustering was applied so that not only users are divided into several clusters based on the
actual rating data and Pearson correlation coefficient but also an incentive/penalty is given to each item according to the preference tendency by users within a same cluster.
As a main result, it was demonstrated that the proposed CBCF method using the IPU model brings a remarkable gain in terms of \textit{recall} or $F_1$ score for a given \textit{precision}. 

A possible direction of future research in this area includes the design of a new clustering-based CF method by exploiting the properties of model-based CF approaches (e.g., matrix factorization).

\section*{Acknowledgement}
The authors would like to thank Younghyun Kim for providing his initial conception of modeling.

\bibliography{RS-IEEEAccess}

\begin{thebibliography}{10}
\providecommand{\url}[1]{#1}
\csname url@samestyle\endcsname
\providecommand{\newblock}{\relax}
\providecommand{\bibinfo}[2]{#2}
\providecommand{\BIBentrySTDinterwordspacing}{\spaceskip=0pt\relax}
\providecommand{\BIBentryALTinterwordstretchfactor}{4}
\providecommand{\BIBentryALTinterwordspacing}{\spaceskip=\fontdimen2\font plus
\BIBentryALTinterwordstretchfactor\fontdimen3\font minus
  \fontdimen4\font\relax}
\providecommand{\BIBforeignlanguage}[2]{{%
\expandafter\ifx\csname l@#1\endcsname\relax
\typeout{** WARNING: IEEEtran.bst: No hyphenation pattern has been}%
\typeout{** loaded for the language `#1'. Using the pattern for}%
\typeout{** the default language instead.}%
\else
\language=\csname l@#1\endcsname
\fi
#2}}
\providecommand{\BIBdecl}{\relax}
\BIBdecl

\bibitem{Adomavicius05}
G.~Adomavicius and A.~Tuzhilin, ``Toward the next generation of recommender
  systems: A survey of the state-of-the-art and possible extensions,''
  \emph{IEEE Trans. Knowl. Data Eng.}, vol.~17, no.~6, pp. 734--749, Jun. 2005.

\bibitem{Linden03}
G.~Linden, B.~Smith, and J.~York, ``Amazon. com recommendations: Item-to-item
  collaborative filtering,'' \emph{IEEE Internet Comput.}, no.~1, pp. 76--80,
  Jan. 2003.

\bibitem{Koren09}
Y.~Koren, R.~Bell, and C.~Volinsky, ``Matrix factorization techniques for
  recommender systems,'' \emph{Computer}, no.~8, pp. 30--37, Aug. 2009.

\bibitem{Konstan12}
J.~A. Konstan and J.~Riedl, ``Recommender systems: from algorithms to user
  experience,'' \emph{User Modeling and User-Adapted Interact.}, vol.~22,
  no.~1, pp. 101--123, Mar. 2012.

\bibitem{Su09}
X.~Su and T.~M. Khoshgoftaar, ``A survey of collaborative filtering
  techniques,'' \emph{Adv. AI}, no.~4, p.~2, Jan. 2009.

\bibitem{Zhou08}
Y.~Zhou, D.~Wilkinson, R.~Schreiber, and R.~Pan, ``Large-scale parallel
  collaborative filtering for the netflix prize,'' in \emph{Proc. 4th Int.
  Conf. Algo. Asp. Inf. Manag. (AAIM '08)}, Shanghai, China, Jun. 2008, pp.
  337--348.

\bibitem{Hu08}
Y.~Hu, Y.~Koren, and C.~Volinsky, ``Collaborative filtering for implicit
  feedback datasets,'' in \emph{Proc. 8th IEEE Int. Conf. Data Mining (ICDM
  '08)}, Pisa, Italy, Dec. 2008, pp. 263--272.

\bibitem{Cai14}
Y.~Cai, H.-F. Leung, Q.~Li, H.~Min, J.~Tang, and J.~Li, ``Typicality-based
  collaborative filtering recommendation,'' \emph{IEEE Trans. Knowl. Data
  Eng.}, vol.~26, no.~3, pp. 766--779, Jan. 2014.

\bibitem{Guo14}
G.~Guo, J.~Zhang, and D.~Thalmann, ``Merging trust in collaborative filtering
  to alleviate data sparsity and cold start,'' \emph{Knowledge-Based Syst.},
  vol.~57, pp. 57--68, Feb. 2014.

\bibitem{Bobadilla12}
J.~Bobadilla, F.~Ortega, A.~Hernando, and J.~Bernal, ``A collaborative
  filtering approach to mitigate the new user cold start problem,''
  \emph{Knowledge-Based Syst.}, vol.~26, pp. 225--238, Feb. 2012.

\bibitem{Sobhanam13}
H.~Sobhanam and A.~K. Mariappan, ``A hybrid approach to solve cold start
  problem in recommender systems using association rules and clustering
  technique,'' \emph{Int. J. Comput. Appl.}, vol.~74, no.~4, pp. 17--23, Jul.
  2013.

\bibitem{Liu14}
H.~Liu, Z.~Hu, A.~Mian, H.~Tian, and X.~Zhu, ``A new user similarity model to
  improve the accuracy of collaborative filtering,'' \emph{Knowledge-Based
  Syst.}, vol.~56, pp. 156--166, 2014.

\bibitem{Huang15}
B.-H. Huang and B.-R. Dai, ``A weighted distance similarity model to improve
  the accuracy of collaborative recommender system,'' in \emph{Proc. 16th IEEE
  Int. Conf. Mobile Data Manag. (MDM)}, Pittsburgh, PA, 2015, pp. 104--109.

\bibitem{Lu12}
Q.~Lu, T.~Chen, W.~Zhang, D.~Yang, and Y.~Yu, ``Serendipitous personalized
  ranking for top-{N} recommendation,'' in \emph{Proc. IEEE/WIC/ACM Int. Conf.
  Web Intell. and Intell. Agent Technol. (WI-IAT '12)}, Washington, DC, Dec.
  2012, pp. 258--265.

\bibitem{Oku11}
K.~Oku and F.~Hattori, ``Fusion-based recommender system for improving
  serendipity,'' in \emph{Proc. ACM Workshop on Novelty Diversity Rec. Sys.
  (DiveRS)}, Chicago, IL, Oct. 2011, pp. 19--26.

\bibitem{Adamopoulos11}
P.~Adamopoulos and A.~Tuzhilin, ``On unexpectedness in recommender systems: Or
  how to expect the unexpected,'' in \emph{Proc. ACM Workshop on Novelty
  Diversity Rec. Sys. (DiveRS)}, Chicago, IL, Oct. 2011, pp. 11--18.

\bibitem{c17}
H.~Andrat and N.~Ansari, ``Analyzing game stickiness using clustering
  techniques,'' \emph{Advances in Comput. and Comput. Sci.}, vol. 554, pp.
  645--654, 2018.

\bibitem{c18}
K.~Chowdhury, D.~Chaudhuri, and A.~K. Pal, ``A novel objective function based
  clustering with optimal number of clusters,'' \emph{Methodologies and
  Application Issues of Contemporary Comput. Framework}, pp. 23--32, 2018.

\bibitem{c19}
S.-H. Lee, Y.-S. Jeong, J.-Y. Kim, and M.~K. Jeong, ``A new clustering validity
  index for arbitrary shape of clusters,'' \emph{Pattern Recognit. Lett.}, vol.
  112, pp. 263--269, Sep. 2018.

\bibitem{c20}
A.~Tehreem, S.~G. Khawaja, A.~M. Khan, M.~U. Akram, and S.~A. Khan,
  ``Multiprocessor architecture for real-time applications using mean shift
  clustering,'' \emph{J. Real-Time Image Process.}, pp. 1--14, 2017.

\bibitem{c21}
T.~VoVan and T.~N. Trang, ``Similar coefficient of cluster for discrete
  elements,'' \emph{Sankhya B}, vol. 80-B, no.~1, pp. 19--36, 2018.

\bibitem{Huang14}
C.-L. Huang, P.-H. Yeh, C.-W. Lin, and D.-C. Wu, ``Utilizing user tag-based
  interests in recommender systems for social resource sharing websites,''
  \emph{Knowledge-Based Syst.}, vol.~56, pp. 86--96, Jan. 2014.

\bibitem{Yin14}
B.~Yin, Y.~Yang, and W.~Liu, ``Exploring social activeness and dynamic interest
  in community-based recommender system,'' in \emph{Proc. 23rd Int. Conf. World
  Wide Web (WWW '14)}, Seoul, Korea, 2014, pp. 771--776.

\bibitem{Guerraoui15}
R.~Guerraoui, A.-M. Kermarrec, R.~Patra, and M.~Taziki, ``{D2P}: distance-based
  differential privacy in recommenders,'' \emph{J. Proc. VLDB Endowment},
  vol.~8, no.~8, pp. 862--873, Apr. 2015.

\bibitem{Koohi16}
H.~Koohi and K.~Kiani, ``User based collaborative filtering using fuzzy
  c-means,'' \emph{Measurement}, vol.~91, pp. 134--139, Sep. 2016.

\bibitem{c26}
J.~Chen, H.~Wang, and Z.~Yan, ``Evolutionary heterogeneous clustering for
  rating prediction based on user collaborative filtering,'' \emph{Swarm and
  Evolutionary Comput.}, vol.~38, pp. 35--41, Feb. 2018.

\bibitem{c27}
U.~Liji, Y.~Chai, and J.~Chen, ``Improved personalized recommendation based on
  user attributes clustering and score matrix filling,'' \emph{Comput.
  Standards \& Interfaces}, vol.~57, pp. 59--67, Mar. 2018.

\bibitem{c28}
J.~Chen, L.~Wei, and L.~Zhang, ``Dynamic evolutionary clustering approach based
  on time weight and latent attributes for collaborative filtering
  recommendation,'' \emph{Chaos, Solitons \& Fractals}, vol. 114, pp. 8--18,
  Sep. 2018.

\bibitem{Wu16}
Y.~Wu, C.~DuBois, A.~X. Zheng, and M.~Ester, ``Collaborative denoising
  auto-encoders for top-{N} recommender systems,'' in \emph{Proc. 9th ACM Int.
  Conf. Web Search and Data Mining (WSDM '16)}, New York, NY, Feb. 2016, pp.
  153--162.

\bibitem{Jiaa15}
C.-X. Jiaa and R.-R. Liua, ``Improve the algorithmic performance of
  collaborative filtering by using the interevent time distribution of human
  behaviors,'' \emph{Phys. A: Stat. Mech. Appl.}, vol. 436, pp. 236--245, Oct.
  2015.

\bibitem{YLiu15}
Y.~Liu, J.~Cheng, C.~Yan, X.~Wu, and F.~Chen, ``Research on the matthews
  correlation coefficients metrics of personalized recommendation algorithm
  evaluation,'' \emph{Int. J. Hybrid Inf. Technol.}, vol.~8, no.~1, pp.
  163--172, Jan. 2015.

\bibitem{Wu15}
D.~Wu, G.~Zhang, and J.~Lu, ``A fuzzy preference tree-based recommender system
  for personalized business-to-business {E}-services,'' \emph{IEEE Trans. Fuzzy
  Syst.}, vol.~23, no.~1, pp. 29--43, Feb. 2015.

\bibitem{Yang16}
Z.~Yang, B.~Wu, K.~Zheng, X.~Wang, and L.~Lei, ``A survey of collaborative
  filtering-based recommender systems for mobile internet applications,''
  \emph{IEEE Access}, vol.~4, pp. 3273--3287, May 2016.

\bibitem{Luxburg07}
U.~Luxburg, ``A tutorial on spectral clustering,'' \emph{Statis. Comput.},
  vol.~17, no.~4, pp. 395--416, Aug. 2007.

\bibitem{fuzzycmeans}
J.~C. Bezdek, E.~Robert, and F.~William, ``{FCM}: The fuzzy $c$-means
  clustering algorithm,'' \emph{Comput. \& Geosci.}, vol.~10, no. 2-3, pp.
  191--203, 1984.

\bibitem{Xu13}
T.~Xu, J.~Tian, and T.~Murata, ``Research on personalized recommendation in
  {E}-commerce service based on data mining,'' in \emph{Proc. Int. MultiConf.
  Eng. Comput. Scient. (IMECS '13)}, Hong Kong, China, Mar. 2013, pp. 313--317.

\bibitem{Pereira15}
A.~L.~V. Pereiraa and E.~R. Hruschkaa, ``Simultaneous co-clustering and
  learning to address the cold start problem in recommender systems,''
  \emph{Knowledge-Based Syst.}, vol.~82, pp. 11--19, Jul. 2015.

\bibitem{c35}
W.~{Hwang}, J.~{Parc}, S.~{Kim}, J.~{Lee}, and D.~{Lee}, ``Told you {I}
  didn’t like it: Exploiting uninteresting items for effective collaborative
  filtering,'' in \emph{Proc. IEEE Int. Conf. Data Eng. (ICDE)}, Helsinki,
  Finland, May 2016, pp. 349--360.

\bibitem{c36}
P.~Cremonesi, Y.~Koren, and R.~Turrin, ``Performance of recommender algorithms
  on top-{N} recommendation tasks,'' in \emph{Proc. ACM Conf. Recommender Syst.
  (RecSys)}, Barcelona, Spain, Sep. 2010, pp. 39--46.

\bibitem{c37}
H.~Steck, ``Training and testing of recommender systems on data missing not at
  random,'' in \emph{Proc. ACM SIGKDD Conf. Knowl. Discovery Data Mining
  (KDD)}, Washington, DC, USA, Jul. 2010, pp. 713--722.

\bibitem{fuzzy}
N.~R. {Pal} and J.~C. {Bezdek}, ``On cluster validity for the fuzzy $c$-means
  model,'' \emph{IEEE Trans. Fuzzy Syst.}, vol.~3, no.~3, pp. 370--379, Aug.
  1995.

\bibitem{coldsetting}
H.~Jazayeriy, S.~Mohammadi, and S.~Shamshirband, ``A fast recommender system
  for cold user using categorized items,'' \emph{Math. Comput. Appl.}, vol.~23,
  no.~1, p.~1, Jan. 2018.

\end{thebibliography}
\bibliographystyle{IEEEtran}

\end{document}